\documentclass[gmd, manuscript]{copernicus}

\usepackage{multirow}
\usepackage{supertabular}
\usepackage{amsthm}
\usepackage{color}

\nolinenumbers
\begin{document}

\title{TRAPPIST-1 Habitable Atmosphere Intercomparison (THAI). \\
 Motivations and protocol version 1.0.}


\Author[1,2,3]{Thomas J.}{Fauchez}
\Author[4,5]{Martin}{Turbet}
\Author[6,7]{Eric T.}{Wolf}
\Author[8]{Ian}{Boutle}
\Author[9,3]{Michael J.}{Way}
\Author[9]{Anthony D.}{Del Genio}
\Author[10]{Nathan J.}{Mayne}
\Author[9,12]{Konstantinos}{Tsigaridis}
\Author[2,3]{Ravi K.}{Kopparapu}
\Author[11]{Jun}{Yang}
\Author[4]{Francois}{Forget}
\Author[2,3]{Avi}{Mandell}
\Author[2,3]{Shawn D.}{Domagal Goldman}

\affil[1]{Goddard Earth Sciences Technology and Research (GESTAR), Universities Space Research Association (USRA), Columbia, Maryland, USA}
\affil[2]{NASA Goddard Space Flight Center, Greenbelt, Maryland, USA}
\affil[3]{GSFC Sellers Exoplanet Environments Collaboration}
\affil[4]{Laboratoire de M\'eteorologie Dynamique, IPSL, Sorbonne Universit\'es, UPMC Univ Paris 06, CNRS, 4 Place Jussieu, 75005 Paris, France}
\affil[5]{Observatoire  Astronomique  de  l’Universit\'e  de  Gen\`eve,  Universit\'e  de  Gen\`eve,  Chemin  des Maillettes 51, 1290 Versoix, Switzerland.}
\affil[6]{Laboratory for Atmospheric and Space Physics, Department of Atmospheric and Oceanic Sciences, University of Colorado Boulder, Boulder, CO, USA}
\affil[7]{NASA Astrobiology Institute’s Virtual Planetary Laboratory, Seattle, WA, USA}
\affil[8]{Met Office, Exeter, UK}
\affil[9]{NASA Goddard Institute for Space Studies, New York, NY 10025, USA}
\affil[10]{University of Exeter, Exeter, UK}
\affil[11]{Dept. of Atmospheric and Oceanic Sciences, School of Physics, Peking University, Beijing, 100871}
\affil[12]{Center for Climate Systems Research, Columbia University, New York, NY, USA.}

\runningtitle{THAI protocol}

\runningauthor{T. Fauchez et al.}

\correspondence{Thomas J. Fauchez (thomas.j.fauchez@nasa.gov)}

\received{}
\pubdiscuss{} 
\revised{}
\accepted{}
\published{}


\firstpage{1}

\maketitle

\begin{abstract}
Upcoming telescopes such as the James Webb Space Telescope (JWST), the European Extremely Large Telescope (E-ELT), the Thirty Meter Telescope (TMT) or the Giant Magellan Telescope (GMT) may soon be able to characterize, through transmission, emission or reflection spectroscopy, the atmospheres of rocky exoplanets orbiting nearby M dwarfs. One of the most promising candidates is the late M dwarf system TRAPPIST-1 which has seven known transiting planets for which  Transit Timing Variation (TTV) measurements suggest that they are terrestrial in nature, with a possible enrichment in volatiles. Among these seven planets, TRAPPIST-1e  seems to be the most promising candidate to have habitable surface conditions, receiving $\sim66\%$ of the Earth's incident radiation, and thus needing only modest greenhouse gas inventories to raise surface temperatures to allow surface liquid water to exist. TRAPPIST-1e is therefore one of the prime targets for JWST atmospheric characterization. In this context, the modeling of its potential atmosphere is an essential step prior to observation. Global Climate Models (GCMs) offer the most detailed way to simulate planetary atmospheres. However, intrinsic differences exist between GCMs which can lead to different climate prediction and thus observability of gas and/or cloud features in transmission and thermal emission spectra. Such differences should preferably be known prior to observations. In this paper we present a protocol to inter-compare planetary GCMs. Four testing cases are considered for TRAPPIST-1e but the methodology is applicable to other rocky exoplanets in the Habitable Zone. The four test cases included two land planets composed with a modern Earth and pure CO$_2$ atmospheres, respectively, and two aqua planets with the same atmospheric compositions. Currently, there are four participating models (LMDG, ROCKE-3D, ExoCAM, UM), however this protocol is intended to let other teams participate as well.

\medskip
\medskip

\end{abstract}


\introduction  
M dwarfs are the most common type of stars in our galaxy and rocky exoplanets orbiting M dwarf stars will likely be the first to be characterized with upcoming astronomical facilities such as the James Webb Space Telescope (JWST). Ultra-cool dwarfs (T < 2700 K) are a sub-stellar class of late M-dwarfs and represent nearly $15\%$ of astronomical objects in the stellar neighborhood of the Sun \citep{Cantrell2013}. Their smaller size compared to other stellar types allows easier detection of rocky exoplanets in close orbits, and this potential was recently realized by the discovery of the TRAPPIST-1 system \citep{Gillon2016, Gillon2017}. Located about 12~pc away TRAPPIST-1 has seven known planets, and is one of the most promising rocky-planet systems for follow-up observations due to the depths of the transit signals \citep{Gillon2017,Luger2017}. Transit Timing Variation (TTVs) measurements of the TRAPPIST-1 planets suggest a terrestrial composition likely enriched in volatiles, and possibly water \citep{Grimm2018}.  Also, it has been found that three planets (TRAPPIST-1 e, f and g) are in the habitable zone \citep[HZ,][]{Kopparapu2013} where surface temperatures would allow surface water to exist \citep{Gillon2017,Wolf2017,Wolf2018,Turbet2018}. 

TRAPPIST-1 is an active M dwarf star \citep{OMalley2017,Wheatley2017,Vida2018} which offers an environment very hostile to the survival of planetary atmospheres. However, \cite{Bolmont2017} and \cite{Bourrier2017} argued that depending on their initial water contents, the TRAPPIST-1 planets could have retained some water presently. Assuming that this water has remained in sufficient quantity, TRAPPIST-1e may be able to maintain habitable conditions (locally or globally around the planet) through a very large set of atmospheric configurations \citep[and references therein]{Wolf2017,Turbet2018,VanGrootel2018}. The first attempt to characterize those planets through transmission spectroscopy has been conducted by \citet{deWit2016,deWit2018} using the Hubble Space Telescope (HST) for the six innermost planets. Their analysis suggests that the TRAPPIST-1 planets do not have a cloud/haze free H$_2$ dominated atmosphere and that a large set of high mean molecular weight atmospheres are possible, such as thick N$_2$, O$_2$, H$_2$O, CO$_2$, or CH$_4$ dominated atmospheres. Using laboratory measurements and models \cite{Moran2018} have also shown that H$_2$ dominated atmospheres with cloud/haze can also be ruled out. Note that the uncertainties of these HST observations were very large, on the order of hundreds of parts per million (ppm) and further investigations with future facilities such as JWST \citep{Barstow2016,Morley2017} will be needed to determine the nature of atmospheres heavier than hydrogen.

Upstream of future JWST characterization of TRAPPIST-1e, it is important to derive constraints on its possible atmosphere to serve as a guideline for the observations. For this purpose, 3-D Global Climate Models (GCMs) are the most advanced tools \citep{Wolf2019}. However, GCMs are very complex models and their outputs can vary from one model to another for a variety of reasons. GCM intercomparisons have been widely used by the Earth science community. For instance the Coupled Model Intercomparison Project (CMIP) initiated in 1995 and currently in its version 6 \citep{Eyring2016}, focuses on the differences in GCM responses to forcings for anthropogenic climate change. While exoplanets receive considerable attention from climate modelers, and atmospheric data from Earth-like worlds may be imminent, to our knowledge only one intercomparison of planetary GCMs has been published \citep{Yang2019}. They found significant differences in global surface temperature between the models for planets around M-dwarf stars due to differences in atmospheric dynamics, clouds  and radiative transfer. However, \cite{Yang2019} concerns planets near the inner edge of the HZ and focuses on highly idealized planetary configurations. Note that another model intercomparison have been run for the exoplanet community: the Palaeoclimate and Terrestrial Exoplanet Radiative Transfer Model Intercomparison Project (PALAEOTRIP). The protocol of this experiment is described in \cite{Goldblatt2017}  and aims to compare a large variety of radiation codes used for paleoclimate or exoplanets sciences, to identify the limit conditions for which each model can produce accurate results. Information and timeline about PALEOTRIP can be found at \url{http://www.palaeotrip.org/}.\\
The motivation behind the TRAPPIST Habitable Atmosphere Intercomparison (THAI), is to highlight  differences among GCM simulations of a confirmed exoplanet, TRAPPIST-1e, that is potentially characterizable in the near term (with JWST or ground-based facilities), and  to evaluate how these differences may impact our interpretations of retrievals of its atmospheric properties from delivered observables. Our objective is also to provide a clear protocol intended for other GCMs to join the intercomparison, which is therefore not only limited to the GCMs presented in this paper. Results of the intercomparison will be presented in a following paper. In this paper, the motivations, including a presentation of TRAPPIST-1e and of the GCMs, are presented in section \ref{motivations}. In section \ref{protocol} we present the THAI protocol describing all the parameters to be set up in the GCM. In Section \ref{outputs}, we list the model parameters to be provided in order for a given model to be comparable to other GCM simulations. A summary is given in section \ref{Outlooks}.

\section{TRAPPIST-1e climate simulation and motivations}\label{motivations}
\subsection{Motivations for a planetary GCM intercomparison}

Global Climate Models (GCMs) are 3-dimensional numerical models designed to represent physical processes at play in planetary atmospheres and surfaces. They are the most sophisticated way to model the atmospheres and oceans of real planets. GCMs can be seen as a complex network of 1-D time-marching climate models connected together through a dynamical core (see description below). Each 1-D column contains physical parameterizations for radiative transfer, convection, boundary layer processes, cloud macroscale and microscale physics, aerosols, precipitation, surface snow and sea ice accumulation, and other processes, at varying levels of complexity.

The motivation behind this experimental protocol is to evaluate how some of the differences between the models can impact the assessment of the planet's habitability and its observables through transmission spectroscopy and thermal phase curves with upcoming observatories such as JWST. The intercomparison protocol was designed to evaluate three possible sources of differences between the models listed below:

\begin{enumerate}
    \item The dynamical core:\\
The dynamical core is a numerical solver of the hydrodynamic equations on the (rotating) planetary sphere. It calculates the winds that transport atmospheric gases, clouds, aerosols, sensible and latent heat, and momentum from one atmospheric column to another.   
    \item The radiative transfer: \\
Each model has its own radiative transfer working assumptions and may use different spectroscopic databases and even different versions of the same spectroscopic database (e.g., HITRAN), collision-induced absorption (CIA), line-by-line versus correlated-k distribution \citep{LacisOinas1991}, line cutoff, spectral resolution, etc. 
    \item The moist physics:\\
The treatment of water in all of its thermodynamic phases is critical for the simulation of habitable planets. In particular cloud and convection process are a significant source of differences between climate models, and these differences are often exacerbated when modeling exoplanets around M-dwarf stars \citep{Yang2014,Yang2019}.

\end{enumerate}

Note that a particular emphasis will be given on the differences of cloud properties between the models because they may have a large impact on the strength of the spectral signatures simulated by current radiative transfer tools \citep{Fauchez2019}. Yet a sufficient understanding of 3D cloud fields is needed to provide realistic observational constraints to observers. It is therefore crucial to address these potential differences between the GCMs.

Four GCMs (in their planetary version) are initially onboard THAI:

\begin{enumerate}
\item the Laboratoire de M\'et\'eorologie Dynamique -  Generic model \citep[LMDG,][a review paper on the model is currently under preparation]{Wordsworth2011},
\item the Resolving Orbital and Climate Keys of Earth and Extraterrestrial Environments with Dynamics \citep[ROCKE-3D, Planet 1.0 version derived from the NASA GISS Model E,][]{Way2017},
\item the Exoplanet Community Atmospheric Model \citep[ExoCAM \footnote{Available on Github, https://github.com/storyofthewolf/ExoCAM\\
Available from NCAR, http://www.cesm.ucar.edu/models/cesm1.2/}, derived from the CAM4 NCAR model,][]{Neale2010},
\item the Met Office Unified Model \citep[UM,][]{Mayne2014,Boutle2017}.
\end{enumerate}
By publishing our protocols in advance of the intercomparison work, we hope that other teams will also use this protocol to compare their own GCM with the four GCMs of this study.

\subsection{The TRAPPIST-1e benchmark}
TRAPPIST-1e is up to now one of the best habitable planet candidates for atmospheric characterization through transmission spectroscopy with JWST. Therefore, it is also an obvious candidate for an experimental protocol for GCM intercomparison.
In Table \ref{tab_TRAPPIST1e} we summarize the TRAPPIST-1e parameters used in the THAI project based on \cite{Grimm2018}.

\begin{table}[H]
\centering
\caption{TRAPPIST-1 stellar spectrum and TRAPPIST-1e planetary parameters from \cite{Grimm2018}}
\begin{tabular}{c c c c c c c}
\hline
\hline
Star $\&$ spectrum & \multicolumn{6}{c}{2600 K BT Settl with Fe/H = 0}\\
Planet & \multicolumn{6}{c}{TRAPPIST-1e}\\
Insolation & \multicolumn{6}{c}{900 $W.m^{-2}$}\\
Rotation period & \multicolumn{6}{c}{6.1 days}\\
Orbital period & \multicolumn{6}{c}{6.1 days}\\
Mass ($M_{\bigoplus}$) & \multicolumn{6}{c}{0.772}\\
Radius ($R_{\bigoplus}$) & \multicolumn{6}{c}{0.910}\\
Density ($\rho_{\bigoplus}$) & \multicolumn{6}{c}{1.024}\\
Gravity ($g_{\bigoplus}$) & \multicolumn{6}{c}{0.930}\\
\hline
\hline
\label{tab_TRAPPIST1e}
\end{tabular}
\end{table}

\section{The THAI Protocol}\label{protocol}
\subsection{Atmospheric configurations}
For THAI, we have chosen a set of four planetary configurations with increasing complexity. We have chosen to start with benchmark cases of dry-land planets with N$_2$- and CO$_2$-dominated atmospheres respectively, which will allow us to assess atmospheric dynamical + boundary layer, and CO$_2$ radiative transfer differences. Next we conduct aquaplanet simulations of N$_2$ and CO$_2$ dominated atmospheres respectively, providing characteristic cold and warm habitable states for TRAPPIST-1e.  By gradually increasing the complexity of our simulations, we hope to be able to parse out meaningful differences between atmospheric dynamical + boundary layer, radiative transfer, and moist physical processes. The motivation for each of these cases is described below:

\begin{itemize}
\item Benchmark case 1 (Ben1): In this case, constituted of 1 bar of N$_2$ only and 400~ppm of CO$_2$, the purpose is to test the differences of the planetary boundary layer (PBL) schemes, the dynamical core and the associated heat redistribution between the different models. Note that N$_2$-N$_2$ CIA should be included.
\item Benchmark case 2 (Ben2): In this case, constituted of 1 bar of CO$_2$, we test the PBL schemes and dynamical core differences as well as the CO$_2$ radiative transfer. 
\item Habitable case 1 (Hab1): In this case, constituted of a modern Earth-like atmosphere of  1~bar of N$_2$ and 400~ppm of CO$_2$, the dynamical core, the clouds and atmospheric processes  are tested together. It is also the most widespread benchmark for habitable planets in the literature \citep{Barstow2016,Morley2017,Lincowski2018}.
\item Habitable case 2 (Hab2): In this case, constituted of 1~bar of CO$_2$, the dynamical core, the  CO$_2$ radiative transfer assumption and the clouds and atmospheric processes  are tested. This case is likely representative of the early Earth (during the Hadean epoch), early Venus, and early Mars, at a time when Martian valley networks and lakes were formed \citep{Haberle2017,Kite2019}.
\end{itemize}

We have therefore two lands planets (Ben1 \& Ben2) and two aqua planets (Hab1 \& Hab2). Note that Ben1 \& Hab1 share the same atmospheric composition of    bar of 1~bar of N$_2$ and 400~ppm of CO$_2$ and Ben2 \& Hab2 share the same atmospheric composition of 1~bar of CO$_2$. The protocol is therefore symmetrical with respect to the atmospheric composition between the land planets and aqua planets. \\

In each case, it is crucial to start each simulation with the same initial conditions. The simplest approach is then to start with an isothermal atmosphere. For THAI, we fixed the initial surface and atmosphere temperature at 300~K. The atmospheric configurations for the two benchmark (dry land) cases and two habitable cases are listed in Table \ref{tab_protocol}, first horizontal block. Note that for Ben2 initial results indicate that some models feature cold trap temperatures on the night-side slightly below the CO$_2$ condensation point at 1~bar (194~K). However, because not all the models include CO$_2$ condensation,   it should be disabled in the models that allow it. Ben2 is thus to be viewed as a idealization for the sake of study. Initial results indicate that Hab1 is representative of a cool, largely ice covered world but with liquid water in the substellar region. Hab2 is significantly warmer than Hab1, owing to a strong CO$_2$ greenhouse effect and the water vapor greenhouse feedback, and is representative of a temperate habitable world. The amount and variability of clouds and the strength of the atmospheric processes should be enhanced providing a more challenging comparison than in Hab1.\\

\begin{figure}[H]
\centering
\resizebox{15cm}{!}{\includegraphics{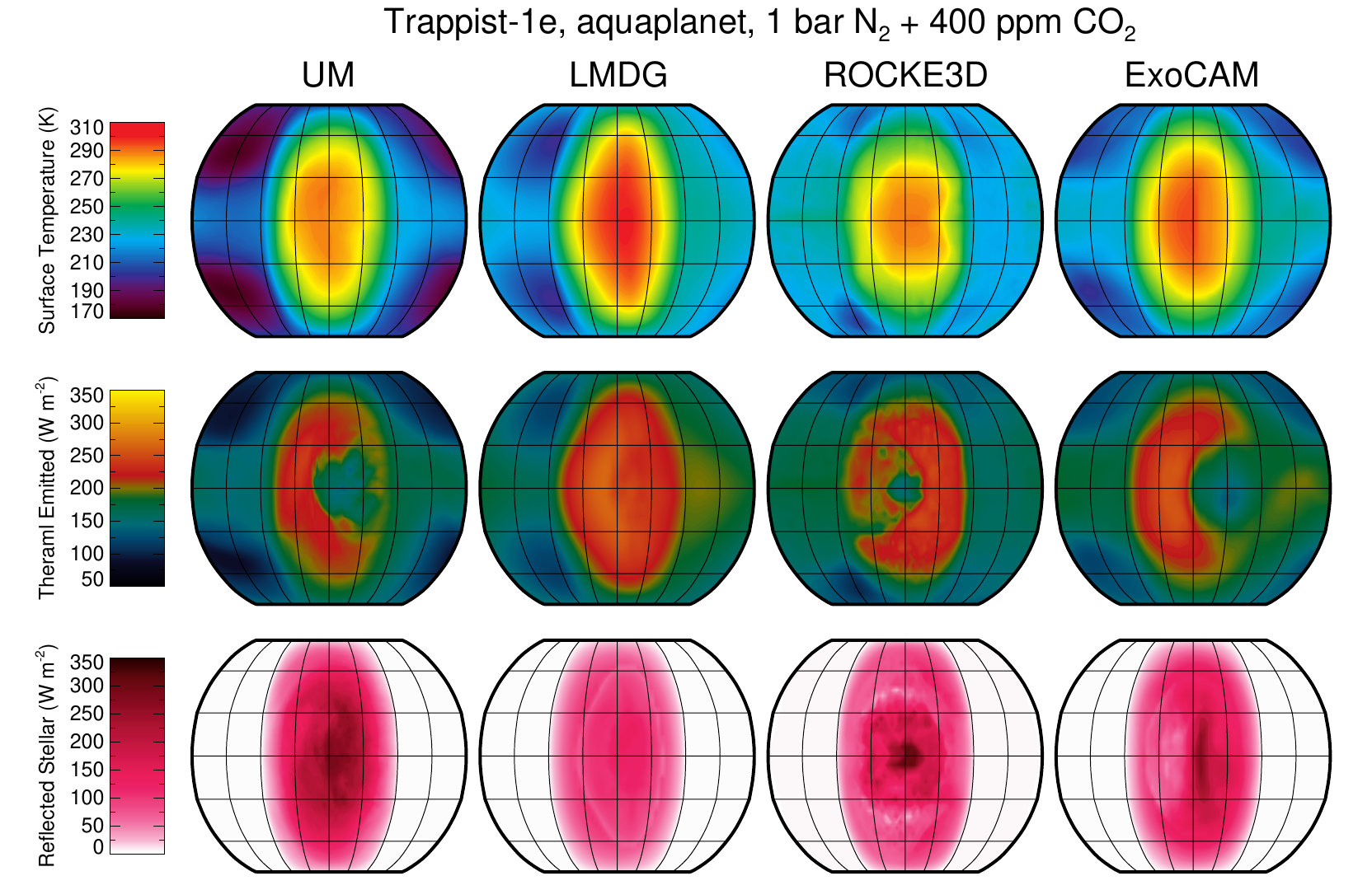}}
\caption{Surface contours for surface temperature, thermal emitted radiation (TOA) and reflected stellar radiation (TOA) for "Hab1" simulated by the four GCMs: the UK Met Office United Model (UM), the Laboratoire M\'et\'eorologie Dynamique Generic model (LMDG), the Resolving Orbital and Climate Keys of Earth and Extraterrestrial Environments with Dynamics (ROCKE-3D), and the National Center for Atmospheric Research Community Atmosphere Model version 4 modified for exoplanets (ExoCAM).}
\label{Fig:prel_results}
\end{figure}

In Fig. \ref{Fig:prel_results} we show results from preliminary simulations on case "Hab1" conducted with four different GCMs; UM, LMDG, ROCKE-3D and ExoCAM.   We show surface contours for surface temperature, thermal emitted radiation (TOA) and reflected stellar radiation (TOA). We can see significant differences in the maximum, minimum and mean values of these parameters between the models. For such a complex atmosphere it is difficult to disentangle the effects leading to these differences. However, it seems clear that the patterns of thermal emitted and reflected stellar TOA fluxes are strongly influenced by the cloud patterns produced by each respective model.  Here we have shown preliminary outputs to demonstrate the feasibility of the described experiments.  In depth analysis of these simulations will be discussed in a following manuscript in preparation.

\subsection{Surface}
The surfaces considered in THAI (Table \ref{tab_protocol}, second horizontal block) are simple. The land planets (Ben1 \& Ben2) are covered by sand with a subsurface depth of at least 3~m with a constant albedo of 0.3. The ocean planets (Hab1 \& Hab2) are fully covered by a 100~m deep slab (no horizontal heat transport) ocean. The ocean albedo is fixed at 0.06 and the ice and snow albedos are fixed at a constant value of 0.25. Note that the sea ice/snow albedo parameterization is a common source of discrepancy between the models. Some models, like ROCKE-3D, account for the spectral dependence of the sea ice albedo over multiple bands and variations due to snowfall, aging, depth and melt ponds while other models, such as LMDG, compute the wavelength-dependent albedo of water ice / snow from a simplified albedo spectral law, calibrated to get an ice / snow bolometric albedo of $\sim0.25$ around an ultra-cool star like TRAPPIST-1 \citep{Joshi2012, vonParis2013,Shields2013}. Differences in sea ice albedo have been found to have a large impact on planetary climate and habitability \citep{Turbet2018}. However, for the sake of this intercomparison, this discrepancy can be easily avoided by fixing the sea ice and snow albedo at a constant bolometric value of $0.25$.

\subsection{Model spatial resolutions and time steps.}
The model spatial resolution is an important parameter because every process taking place at a sub-grid level would be parameterized and those parameterizations often diverge between the models. Similarly the model time steps control the numerical stability and accuracy. However,  the choices for those are fundamental to how each model operates under a given parameterization and arbitrary fixing these parameters may prevent some model to correctly and fairly perform the intercomparison. In addition, models should be compared using the specifications  that they commonly use for exoplanet studies. Therefore, for the sake of the THAI, we do not impose the model spatial resolution nor time steps. Note that we however recommend (but this is not a requirement) the radiative time step (a parameter much more flexible than the others among the models) to be set up at 1800~s. This value should provide a good coupling of the radiation with temporal changes to the atmosphere without slowing down too much the model.

We also ask the contributing scientists to disable the sub-grid gravity wave parameterizations in their model. Indeed, all the models do not have implemented a gravity wave parameterization and some have prescribed or predicted gravity wave formation, tuned for Earth topography and meteorology. Therefore, to avoid differences in atmospheric dynamics especially above the tropopause, we recommend to not include the sub-grid gravity wave parameterizations in this intercomparison. Gravity waves whose wavelengths are greater than the model grid are explicitly resolved in the models and do not need to be modified. \\

Note that under the requirements of the protocol, the atmospheric simulation of TRAPPIST-1e may actually not represent what each individual model can simulate with all their parameterizations fully activated. This is especially true for the sea ice and snow albedo parameterization. Therefore, complementary to the Hab1 case, we propose the Hab1* which should be simulated with the commonly used model parameterizations fully activated. Therefore, only the requirements on the atmospheric composition (1~bar of N$_2$ and 400~ppm of CO$_2$) and the planet and star properties of Table \ref{tab_TRAPPIST1e} are constrained for Hab1*.

\begin{table}[H]
\centering
\caption{THAI experimental protocol.}

\begin{tabular}{c c c| c c }
\hline
\hline
Case & Ben1 & Ben2 & Hab1 & Hab2 \\
\hline
& \multicolumn{4}{c}{\bf{Atmospheres}}\\
Composition & 1 bar N$_2$ + &  1 bar CO$_2$ & 1 bar N$_2$ +  &  1 bar CO$_2$ \\
& 400 ppm CO$_2$ &     &  400 ppm CO$_2$ &  \\

Molecular air mass (dry) & 28 & 44 &  28  & 44\\
Initial state & \multicolumn{2}{c|}{Isothermal 300~K} & \multicolumn{2}{c}{Isothermal 300~K}\\
\hline
& \multicolumn{4}{c}{\bf{Surfaces}}\\
Type & \multicolumn{2}{c|}{Land only} & \multicolumn{2}{c}{Ocean planet}\\
Composition & \multicolumn{2}{c|}{Sand} & \multicolumn{2}{c}{Slab ocean}\\
Albedo & \multicolumn{2}{c|}{0.3}  & \multicolumn{2}{c}{Liquid water: 0.06} \\
 & & & \multicolumn{2}{c}{Ice/snow: 0.25} \\
Heat capacity (J/m$^3$/K) & \multicolumn{2}{c|}{$2\cdot10^6$} & \multicolumn{2}{c}{$4\cdot10^6$ } \\
Thermal inertia (J/m$^2$/K/s$^2$) & \multicolumn{2}{c|}{2000} & \multicolumn{2}{c}{12000} \\
Momentum roughness length (m) &\multicolumn{2}{c|}{0.01} & \multicolumn{2}{c}{0.01}\\
Heat roughness length (m) & \multicolumn{2}{c|}{0.001} & \multicolumn{2}{c}{0.001}\\
Depth of the subsurface / ocean (m) &  \multicolumn{2}{c|}{> 3} & \multicolumn{2}{c}{100}\\
\hline
Cautions: & \multicolumn{4}{c}{disable sub-grid gravity wave parameterization} \\
& \multicolumn{4}{c}{disable CO$_2$ condensation} \\
\hline
\hline
\label{tab_protocol}
\end{tabular}
\end{table}

\section{Outputs}\label{outputs}
\begin{figure}[H]
\centering
\resizebox{15cm}{!}{\includegraphics{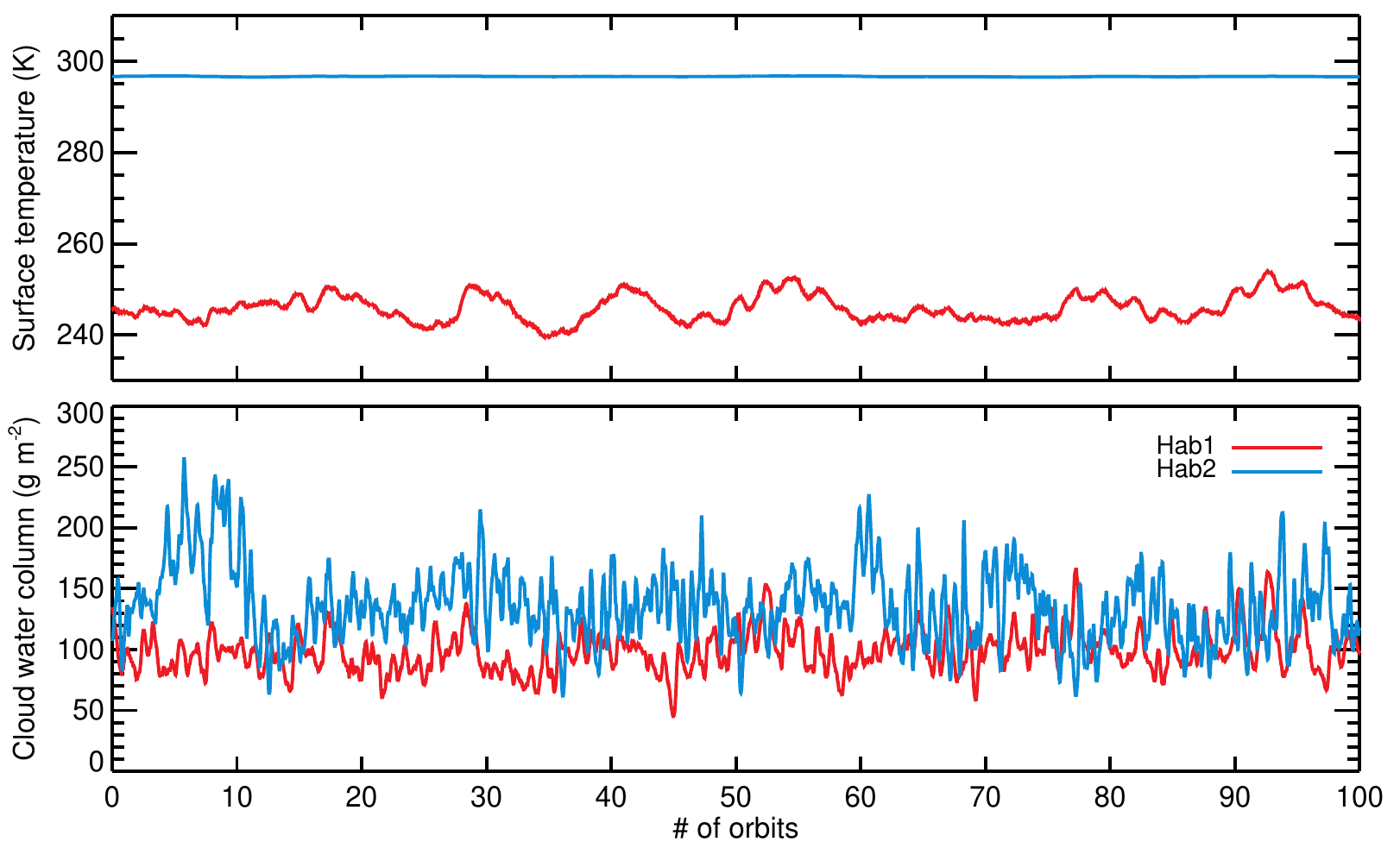}}
\caption{Globally averaged surface temperature (top panel) and cloud water column (bottom panel) as a function of the number of orbits for Hab1 \& Hab2 simulated with ExoCAM \citep{Wolf2015}. Surface temperature for Hab1 and cloud water column for the 2 cases vary by a couple of tens of percents on a timescale of 10 orbits due to weather patterns.}
\label{fig:Hab1Hab2}
\end{figure}

 To compare the difference between models of a particular (instantaneous) output variable, both the average and standard deviation over the specified frequency and number of orbits for the case will be computed.
Four categories of outputs frequently used in climate simulations have been selected: radiation, surface, atmospheric profiles and clouds. The radiation outputs are the outgoing longwave radiation (OLR) and absorbed shortwave radiation (ASR) for clear and cloudy skies, also commonly known as emitted thermal and absorbed stellar fluxes, respectively, both at the top of the atmosphere (TOA). The surface outputs are the temperature map, the downward total SW flux and net LW flux and the open ocean fraction (for Hab1/Hab1* \& Hab2 only). The atmosphere outputs are the temperature and the U, V and W wind speed profiles. Finally, the cloud outputs For Hab1/Hab1* \& Hab2 are the water vapour and cloud condensed water and ice integrated columns, and the cloud profiles of the cloud fraction and the mass mixing ratio for the liquid, ice and both combined.
Also in these two cases, the spatial and temporal variability is much weaker than in Hab1/Hab1* \& Hab2. Therefore, to mitigate the amount of data we choose to only output data for ten consecutive orbits (with a 6~hour output frequency). Concerning Hab1/Hab1* \& Hab2, we can see in Figure \ref{fig:Hab1Hab2} that weather patterns modulate the surface temperature and cloud water column of Hab1 on a period nearly equal to 10 orbits. Also Hab2 (hotter than Hab1) has a more efficient heat transport and is therefore more homogeneous in temperature but the cloud variability is very important. Therefore, more orbits (100) are needed in order to smooth out this variability.  A summary of the output parameters is given in Table \ref{tab_outputs}.

All the simulations should have reached radiative equilibrium at TOA at $\sim\pm 1\ W\cdot m^{-2}$. If such limit can't be achieved by the model, the radiative equilibrium can be established if no discernible trend are observable in the last 10 year average global mean temperature.\\

To facilitate comparison between each GCM, we ask the contributing scientists to provide their outputs in netCDF format.  The contributing scientist will be able to upload their data on a public permanent repository at \url{https://thai.emac.gsfc.nasa.gov/dataset/thai} after requesting an IP address authorization to Thomas Fauchez (thomas.j.fauchez@nasa.gov).

\begin{table}[H]
\centering
\caption{Instantaneous fields to be output by the GCM. For each diagnostic, the mean value and the standard deviation are computed from data output at the specified frequency and number of orbits for the case. OLR and ASR correspond to outgoing longwave radiation (at TOA) and absorbed shortwave radiation (at TOA), respectively, SW and LW correspond to shortwave and longwave, respectively, CF corresponds to cloud fraction and MMR at mass mixing ratio.}
\begin{tabular}{p{3cm} p{3cm} p{3cm} p{4cm} p{3cm}}
\hline
\hline
Case & Ben1 & Ben2 & Hab1/Hab1* & Hab2\\
\hline
Number of orbits & 10 & 10 & 100 & 100 \\
Frequency (hours) & 6 & 6 & 6 & 6\\
\hline
 \multicolumn{5}{c}{\textbf{2D maps}}\\
\hline
Radiation &\multicolumn{4}{c}{OLR (clear/cloudy)}\\
& \multicolumn{4}{c}{ASR (clear/cloudy)}\\
\hline
Surface & \multicolumn{4}{c}{temperature map}\\
& \multicolumn{4}{c}{downward total SW flux}\\
& \multicolumn{4}{c}{Net LW flux}\\
& $\varnothing$  & $\varnothing$ & \multicolumn{2}{c}{open ocean fraction}\\
\hline
Clouds &$\varnothing$  & $\varnothing$ & \multicolumn{2}{c}{total/liquid/ice/vapor column}\\
\hline
 \multicolumn{5}{c}{\textbf{Vertical profiles}}\\
 \hline
Atmospheric  & \multicolumn{4}{c}{temperature}\\
profiles & \multicolumn{4}{c}{U, V , W wind speed}\\
 & \multicolumn{4}{c}{heating rates (SW/LW)}\\
& $\varnothing$ & $\varnothing$ & \multicolumn{2}{c}{specific + relative humidity}\\
\hline
Cloud profiles   & $\varnothing$ & $\varnothing$ & \multicolumn{2}{c}{CF (total/liquid/ice) [\%]}\\
 &$\varnothing$ & $\varnothing$ & \multicolumn{2}{c}{MMR (total/liquid/ice) [kg/kg]}\\

\hline
\hline
\label{tab_outputs}
\end{tabular}
\end{table}

The main objective of THAI is to highlight how differences in atmospheric profiles produced by each GCM are going to impact the predictions of atmosphere detectability and observational constraints for habitable planet targets such as TRAPPIST-1e \citep{Morley2017,Fauchez2019}. Therefore, in addition to the parameters of Table \ref{tab_outputs},  we will emphasize the differences between the models in term of the planet's climate and habitability with a particular attention on the cloud coverage. Also, the objective will be to identify and quantify the differences on the simulated JWST observations, through simulated transmission spectra (in NIRSpec prism and MIRI LRS ranges) and thermal phase curves (in MIRI LRS range) due to the differences of atmospheric profiles (temperature, pressure and gas mixing ratios) output by each GCM. The planetary spectrum generator (PSG, \cite{Villanueva2018}) will be used to simulate transmission and emission spectra. The comparison of the spectra for Hab1 \& Hab2 cases will therefore highlight the sensitivity of model characteristics to predict transmission spectra of habitable planets. \\

Note that while additional simulations with a simple Newton cooling model, a 1-D column model, or with cloud radiative effects disabled would help to better understand the differences due to the dynamical cores and cloud physics, they would also dramatically increase the computationnal time, amount of data and effort. THAI aims to be easily reproducible and not time consuming in order to reach many GCM user groups. The five simulations propose in THAI should be enough to understand the main differences between the GCMs and their impact on the observables. THAI could also be used as a benchmark for future GCM intercomparisons that specifically aim to understand each differences between the models.

\section{Summary}\label{Outlooks} 
THAI is an intercomparison project of planetary GCMs focused on the exciting new habitable planet candidate, TRAPPIST-1e. Because rocky exoplanets in the Habitable Zone of nearby M dwarfs have the highest chance to be the first Earth-size exoplanets to be characterized with future observatories, TRAPPIST-1e is currently the best benchmark we could think of to compare the capability of planetary GCMs. In this first paper we have presented the planet and GCM parameters to be used in this experiment which already has four GCMs onboard (LMDG, ROCKE-3D, ExoCAM and UM), but we hope more GCMs will join the project. The results of the comparison of these four models will be given in a second paper and a THAI workshop is planned for fall 2020.

\codeavailability{ExoCAM \citep{Wolf2015} is available on Github, https://github.com/storyofthewolf/ExoCAM. The Met Office Unified Model is available for use under licence, see http://www.metoffice.gov.uk/research/modelling-systems/unified-model. ROCKE-3D is public domain software and available for download for free from \url{https://simplex.giss.nasa.gov/gcm/ROCKE-3D/}. Annual tutorials for new users take place annually, whose recordings are freely available on line at \url{https://www.youtube.com/user/NASAGISStv/playlists?view=50&sort=dd&shelf_id=15}. LMD-G is obtainable upon request from Martin Turbet (martin.turbet@lmd.jussieu.fr) and François Forget (francois.forget@lmd.jussieu.fr).} 

\dataavailability{No data has been used.} 












\competinginterests{No competing interests are present.} 

\paragraph*{Author contribution}
T.J.F. lead the THAI project and has written the manuscript. E.T.W ran the simulation for Fig 1. and plotted the figures.
Every author contributed to the development of the THAI protocol, to the discussions and to the editing of the manuscript.

\begin{acknowledgements}
Goddard affiliates are thankful for support from GSFC Sellers Exoplanet Environments Collaboration (SEEC), which is funded by the NASA Planetary Science Divisions Internal Scientist Funding Model. M.T. acknowledges the use of the computing resources on OCCIGEN (CINES, French National HPC).\\
This project has received funding from the European Union’s Horizon 2020 research and innovation program under the Marie Sklodowska-Curie Grant Agreement No. 832738/ESCAPE.  M.W. and A.D. acknowledge funding from the NASA Astrobiology Program through participation in the Nexus for Exoplanet System Science (NExSS).\\
The THAI GCM intercomparison team is grateful to the Anong's THAI Cuisine restaurant in Laramie for hosting its first meeting on November 15, 2017.
We would like to thank the reviewer Daniel D.B. Koll, the anonymous reviewer and the topical editor Julia Hargreaves  for comments that greatly improved our manuscript.
\end{acknowledgements}







 \bibliographystyle{copernicus}
 \bibliography{ref_latex.bib}

\end{document}